\documentclass[%
 reprint,
superscriptaddress,
 amsmath,amssymb,
 aps,
 prl,
]{revtex4-1}

\usepackage{graphicx}
\usepackage{epstopdf}
\usepackage{xcolor}
\usepackage{dcolumn}
\usepackage{bm}
\usepackage{hyperref}

\begin{document}

\preprint{APS/123-QED}

\title{Direct frequency-comb-driven Raman transitions in the terahertz range}


\author{C. Solaro}
\email{solaro@phys.au.dk}
\author{S. Meyer}
\author{K. Fisher}
\author{M. V. DePalatis}
\author{M. Drewsen}
\email{drewsen@phys.au.dk}
\affiliation{Department of Physics and Astronomy, Aarhus University, DK-8000 Aarhus C, Denmark}

\date{\today}

\begin{abstract}
We demonstrate the use of a femtosecond frequency comb to coherently drive stimulated Raman transitions between terahertz-spaced atomic energy levels. More specifically, we address the $3d~^2D_{3/2}$ and $3d~^2D_{5/2}$ fine structure levels of a single trapped $^{40}$Ca$^+$ ion and spectroscopically resolve the transition frequency to be \mbox{$\nu_D = 1{,}819{,}599{,}021{,}534 \pm 8$ Hz}. The achieved accuracy is nearly a factor of five better than the previous best Raman spectroscopy, and is currently limited by the stability of our atomic clock reference. Furthermore, the population dynamics of frequency-comb-driven Raman transitions can be fully predicted from the spectral properties of the frequency comb, and Rabi oscillations with a contrast of \mbox{99.3(6)\%} and millisecond coherence time have been achieved. Importantly, the technique can be easily generalized to transitions in the sub-kHz to tens of THz range and should be applicable for driving, e.g., spin-resolved rovibrational transitions in molecules and hyperfine transitions in highly charged ions.


\end{abstract}

\maketitle

It is difficult to overestimate the impact of femtosecond frequency combs \cite{Diddams2000} on optical frequency metrology \cite{Haensch2006,Hall2006} as well as on high resolution spectroscopy \cite{Stowe2008a}. Such combs can be used to directly induce transitions not only in the visible spectrum for high precision measurements \cite{Marian2004, Wolf2009}, but also in regions not easily accessible with CW lasers, for spectroscopy (e.g. in the deep-ultraviolet \cite{Witte2005}) as well as for cooling and trapping of exotic atomic species \cite{Jayich2016,Kielpinski2006}. More notably, their large bandwidth and tunability allow one to address multiple transitions, making the optical frequency comb a promising tool for manipulating complex multilevel quantum systems. In this context, frequency combs have already been applied to the cooling of \cite{Viteau2008,Lien2014} and proposed for the quantum control of \cite{Peer2007,Shapiro2008} molecular rovibrational states. In particular, deterministic and coherent manipulation of pure rotational states in molecules could also be achieved by driving stimulated Raman transitions with femtosecond frequency combs, hence exploiting their high tunability in the terahertz (THz) range \cite{Ding2012, Leibfried2012a}. Previously, pulse trains \cite{Mlynek1981} or mode-locked picosecond lasers \cite{Fukuda1981} have been applied in the MHz to the few GHz range in atomic ensembles, and more recently, Hayes et al. \cite{Hayes2010} used a picosecond frequency comb to coherently manipulate hyperfine levels separated by $\sim 12.6$ GHz in trapped and laser-cooled $^{171}$Yb$^+$ ions.

\begin{figure}[t]
\includegraphics[width=0.48\textwidth]{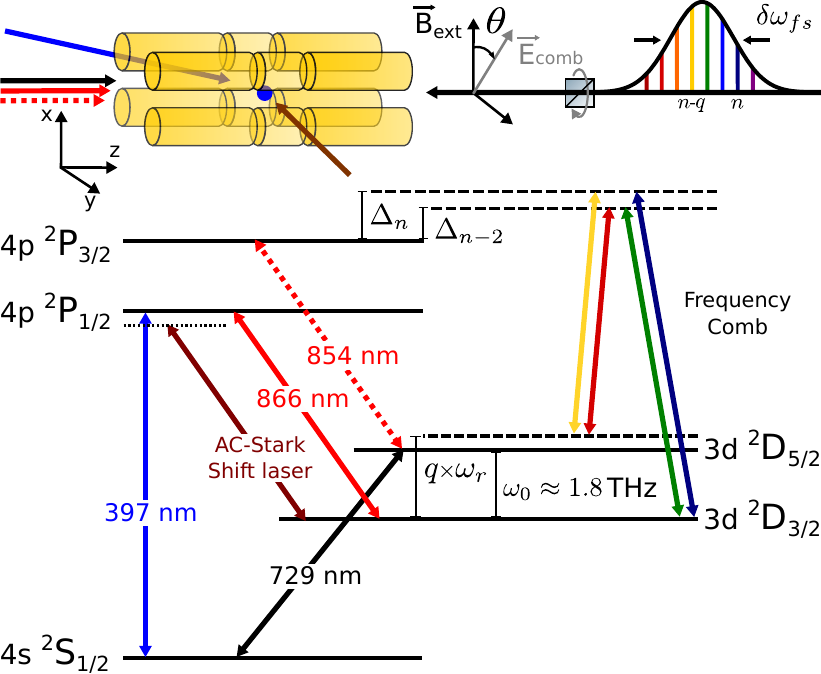}
\caption{Schematic of the experimental setup and the relevant electronic levels of the Ca$^+$ ion. The 397-nm beam, propagating along the $\frac{1}{\sqrt{2}}(\hat{y}+\hat{z})$ direction, is $\pi$-polarized with respect to the external magnetic field $\vec{B}_{ext}$. The 866-nm, 854-nm and 729-nm beams, propagating along $\hat{z}$, have a linear polarization rotated by 45$^{\circ}$ with respect to $\vec{B}_{ext}$. The AC-Stark-shift laser beam, propagating along $\hat{y}$, has a linear polarization rotated to maximize the AC-Stark shift on the selected Zeeman sub-levels of the $3d~^2D_{3/2}$ state. The frequency comb beam, propagates along $\hat{z}$ with a well-defined linear polarization $\vec{E}_{comb}$ set by a rotatable Glan-Taylor polarizer.}
\label{Fig1}
\end{figure}

In this Letter, we report on expanding the comb technique to the THz range by using a femtosecond frequency comb to drive stimulated Raman transitions between the $3d~^2D_{3/2}$ and $3d~^2D_{5/2}$ fine structure states of a single trapped $^{40}$Ca$^+$ ion. Precise control of the mode-locked laser's repetition rate $\omega_r$ allows us to drive transitions between selected Zeeman sub-levels. Whenever a harmonic of the repetition rate is equal to the energy splitting ($q\times\omega_r = \omega_{0}$), and if the comb's spectral bandwidth $\delta\omega_{fs}$ is on the order of $\omega_{0}$, all frequency components of the comb contribute to drive the Raman transition. Interestingly, if $\delta\omega_{fs}\gg\omega_{0}$, each comb tooth contributes twice, and the resulting Raman transition rate is twice that of two phase-locked CW lasers with identical and combined intensity equal to the comb intensity \cite{[{See Supplemental Material at [URL will be inserted by publisher] for details regarding the experiment and our model, which includes Ref. \cite{James1998}}] supmat}. Since this process relies only on the frequency difference between Raman pairs, the carrier envelope offset frequency does not need to be locked. However, because the effective comb's linewidth is proportional to $q$, expanding the technique from GHz to THz transitions requires better stabilization of the repetition rate. In addition, non-linear dispersion, which increases with the comb's spectral bandwidth, can undermine the phase relation between pairs of comb teeth and hence influence the effective Raman coupling strength.

An overview of our experimental setup is presented in figure \ref{Fig1}. We use a linear Paul trap, detailed in Ref. \cite{Drewsen2003}, consisting of four rods, each sectioned into three electrodes. By applying suitable AC and DC voltages, an effective 3D harmonic confining potential is created with axial and radial frequencies of $\{\omega_z,\omega_r\} = 2\pi\times \{509, 730\}$ kHz respectively. A single $^{40}$Ca$^+$ ion is loaded into the trap via isotope-selective photoionization of atoms in a neutral calcium beam \cite{Kjergaard2000,Mortensen2004}. An external magnetic field of 6.500(3) G along the $x$-direction lifts the Zeeman degeneracy of the involved electronic energy levels by a few MHz, defining a natural quantization axis. An experimental cycle starts with Doppler cooling on the $S_{1/2} \leftrightarrow P_{1/2}$ transition, carried out using a single 397-nm laser beam. Simultaneously, repump beams at both 866~nm and 854~nm clear population out of the $D_{3/2}$ and $D_{5/2}$ states, respectively. Next, the ion is sideband-cooled to the ground state of the trapping potential along the $z$-direction using a narrow-linewidth 729-nm laser beam addressing the $S_{1/2} \leftrightarrow D_{5/2}$ electric-quadrupole transition.  The sideband cooling sequence, detailed in \cite{Poulsen2012a}, is followed by optical pumping into one of the two $|S_{1/2},m_j=\pm1/2\rangle$ states. Initialization to the chosen $|D_{5/2},\pm m_j\rangle$ state with day-to-day efficiency $\geq97\%$ is then achieved by rapid adiabatic passage (RAP) \cite{Turrin1977,Wunderlich2007} from the $|S_{1/2},\pm1/2\rangle$ state. After this preparation sequence, we drive Raman transitions between the $|D_{5/2},m_j\rangle$ and $|D_{3/2},m_j'\rangle$ states with a femtosecond frequency comb laser. The state of the ion is read out by the electron-shelving technique \cite{Dehmelt1982} through addressing the $S_{1/2} \leftrightarrow P_{1/2}$ and $D_{3/2} \leftrightarrow P_{1/2}$ transitions. This cycle is repeated 100 times to extract the mean transition probability. 

\begin{figure}[t]
\includegraphics[width=0.48\textwidth]{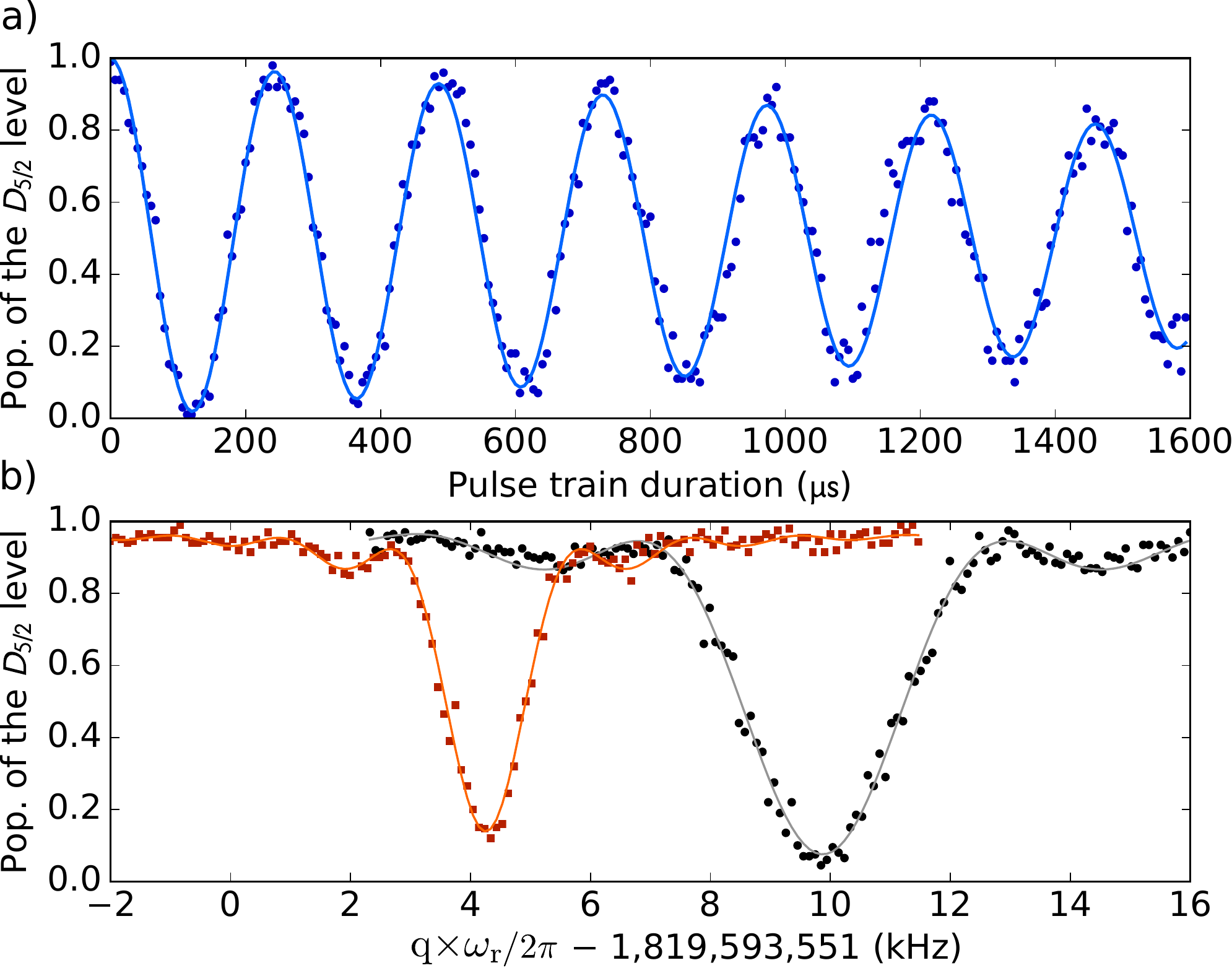}
\caption{(a) Evolution of the population of the $D_{5/2}$ state versus duration of the frequency comb's pulse train. Points correspond to the experimental data, while the solid line is a fit to the data with 99.3(6)\% contrast, an exponential decay time of 3.2(2) ms and a Rabi frequency of \mbox{$\Omega_R=2\pi\times 4.121(3)$ kHz}. Note that the contrast, depending also on state initialization, is a lower estimate to the Raman transfer efficiency. (b) Typical spectra of the $|J,m_j\rangle : |\frac{5}{2},-\frac{3}{2}\rangle \rightarrow |\frac{3}{2},-\frac{3}{2}\rangle$ transition performed at two different comb's intensities and with $\vec{E}_{comb}\parallel\vec{B}_{ext}$. The measured spectra are shifted from the expected transition frequency (calculated by taking the linear Zeeman shift into account) by $\delta\nu^{AC}_D=9.873(12)$ kHz (black points) and $\delta\nu^{AC}_D=4.253(8)$ kHz (red squares) due to the comb's AC-Stark shift induced by off-resonant coupling to the $P_{1/2,3/2}$ states. Solid curves are sinc$^2$ fits to the data.}
\label{Fig2}
\end{figure}

\begin{figure}[ht]
\includegraphics[width=0.48\textwidth]{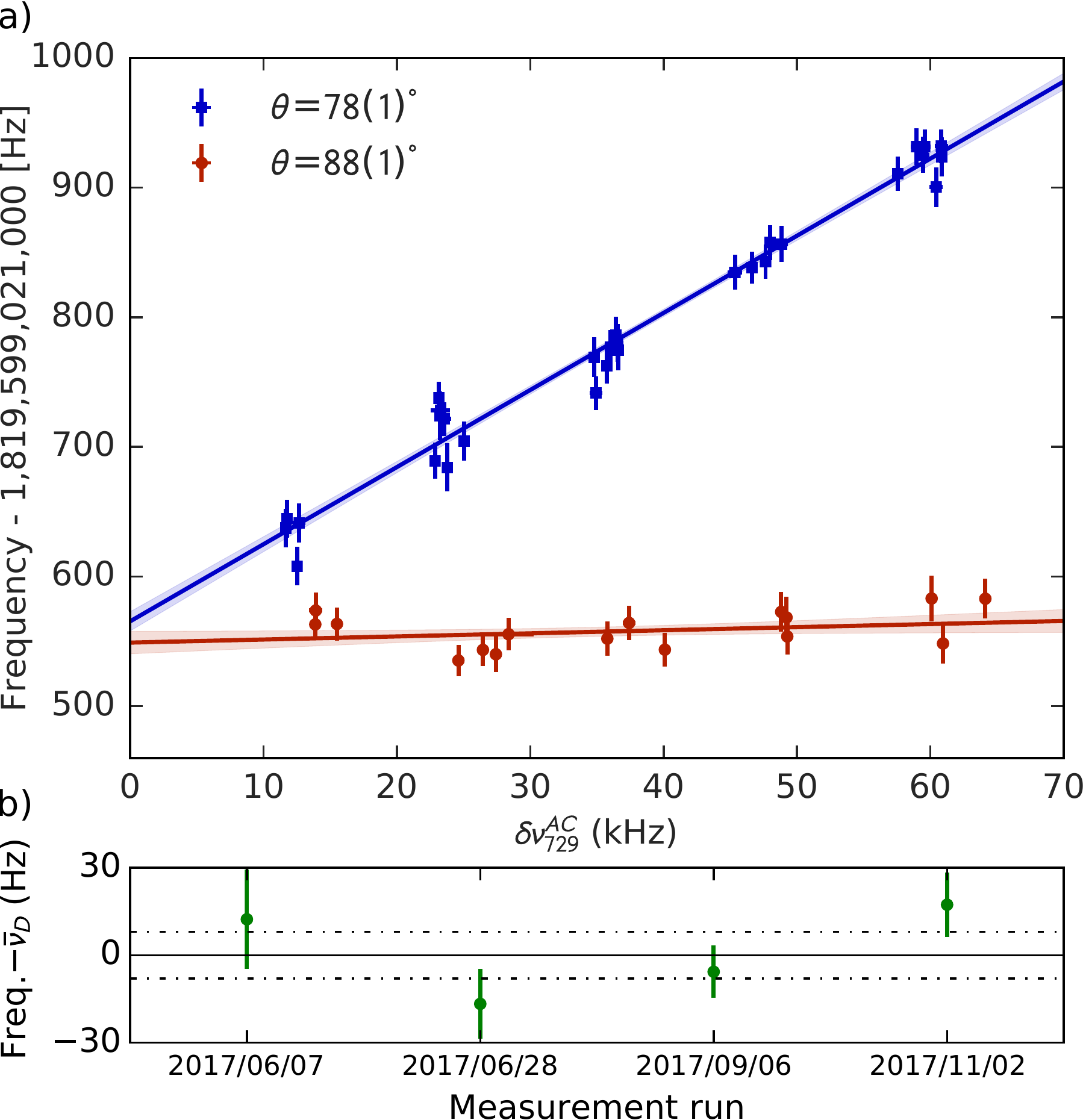}
\caption{ (a) Measured transition frequency versus the differential AC-Stark shift induced by the frequency comb on the $|S_{1/2},-\frac{1}{2}\rangle \leftrightarrow |D_{5/2},-\frac{5}{2}\rangle$ transition ($\delta\nu^{AC}_{729}\propto I$). Blue squares (red points) correspond to a comb polarization rotated by $\theta=78(1)^{\circ}$ ($\theta=88(1)^{\circ}$) from the quantization axis. Solid lines show the linear fit to the data used to extrapolate the fine structure level separation to zero intensity. The measured frequency is extracted from the average of the resonance frequencies of the $|\frac{5}{2},\frac{1}{2}\rangle \rightarrow |\frac{3}{2},\frac{1}{2}\rangle$ and $|\frac{5}{2},-\frac{1}{2}\rangle \rightarrow |\frac{3}{2},-\frac{1}{2}\rangle$ transitions. The spread along the $x$-axis is mainly due to pointing instabilities of the frequency comb beam. (b) Absolute frequency measurements on different dates used to determine the fine structure level separation \mbox{$\bar{\nu}_D = 1{,}819{,}599{,}021{,}555 \pm 8$ Hz}. The dashed lines correspond to the 8 Hz standard deviation of the mean.}
\label{Fig3}
\end{figure}

The femtosecond frequency comb laser is a commercial mode-locked fiber laser from MenloSystems (model FC1500-250-WG \cite{menlo}) with a carrier frequency of $\omega_c/2\pi \approx 380$~THz, blue-detuned on average by $\Delta/2\pi\approx 29$~THz from the $D_{5/2} \leftrightarrow P_{3/2}$ transition, and spectrally truncated at a minimum detuning of 7 THz by two short-pass filters (Semrock FF01-842/SP-25). The repetition rate of this laser is $\omega_r/2\pi = 250$~MHz. It can be finely adjusted (by a few kHz) in between measurements to address transitions between different Zeeman sub-levels. The frequency comb beam $1/e^2$ radius at the position of the ion is 34(2) $\mu$m, and the average power ranges from 18 to 90 mW. The beam is blocked by a mechanical shutter during preparation and readout. For fast effective shuttering of the comb beam, an AOM-controlled AC-Stark-shift laser, detuned by 2 GHz from the $D_{3/2} \leftrightarrow P_{1/2}$ transition, is used to shift the $D_{3/2}$ level out of resonance by about 50 kHz with a switching time of about $150$ ns. This allows for precise coherent manipulation of the populations of the $D_{3/2}$ and $D_{5/2}$ states with a time resolution much smaller than the ms-timescale Raman Rabi oscillations shown in Fig. \ref{Fig2}(a). Damping of the Raman Rabi oscillations is due to both magnetic field fluctuations and instabilities of the comb's repetition rate. The 1/e decay time extracted from an exponential decay fit to the data is $T_{coh} = 3.2(2)$ ms. For $\pi$-pulses shorter than $T_{coh}$, the transition lineshape is a Fourier-limited sinc$^2$ and data can be fit to extract the transition frequency within 10 to 20 Hz uncertainty as exemplified by Fig.~\ref{Fig2}(b). The 6 kHz frequency offset between the two spectra shown in figure~\ref{Fig2}(b) is due to the differential AC-Stark shift $\delta\nu^{AC}_D$ induced by the comb resulting from off-resonant coupling of the $D_{3/2,\ 5/2}$ states to the $P_{1/2,\ 3/2}$ states. The unshifted transition frequency is obtained from extrapolating the measured frequencies to zero light intensity. To determine the laser intensity at the position of the ion, the differential AC-Stark shift $\delta\nu^{AC}_{729}$ induced by the comb on the $|S_{1/2},-\frac{1}{2}\rangle \leftrightarrow |D_{5/2},-\frac{5}{2}\rangle$ transition is measured with the 729-nm laser between each Raman scan. Doing so reduces the error on the extrapolated frequency due to, e.g., pointing instabilities. In addition, we take advantage of the existence of (for some transitions) a ``magic polarization'' for which $\delta\nu^{AC}_D=0$. This is for instance the case for the $|\frac{5}{2},\pm\frac{3}{2}\rangle \rightarrow |\frac{3}{2},\pm\frac{3}{2}\rangle$ as well as the $|\frac{5}{2},\pm\frac{1}{2}\rangle \rightarrow |\frac{3}{2},\pm\frac{1}{2}\rangle$ transitions \cite{supmat}. The other dominant systematic shift is the first-order differential Zeeman shift $\delta\nu^{Z}_D$ induced by the static magnetic field defining the quantization axis. This shift is about three times smaller for the $|\frac{5}{2},\pm\frac{1}{2}\rangle \rightarrow |\frac{3}{2},\pm\frac{1}{2}\rangle$ transitions than for the $|\frac{5}{2},\pm\frac{3}{2}\rangle \rightarrow |\frac{3}{2},\pm\frac{3}{2}\rangle$ transitions and cancels out when averaging a pair of symmetric transitions. To eliminate this shift, the two $|\frac{5}{2},\pm\frac{1}{2}\rangle \rightarrow |\frac{3}{2},\pm\frac{1}{2}\rangle$ transitions are probed in the experiments. With typically 30 points taken interleaved on each transition within 15 min, the measured fine structure level separation at a given comb intensity is determined with 20 Hz resolution. This sequence is then repeated by alternating measurements at five different comb intensities to extrapolate the AC-Stark-shifted transition frequency to zero intensity. Figure \ref{Fig3}(a) shows the measurements for two different comb polarizations $\vec{E}_{comb}$. Blue squares correspond to a linear polarization rotated by $\theta=78(1)^{\circ}$ with respect to the quantization axis (see Fig. \ref{Fig1}). Red points correspond to the magic polarization ($\theta=88(1)^{\circ}$) and show essentially no dependence with respect to $\delta\nu^{AC}_{729}$ (i.e. the comb intensity). From this measurement, the extrapolated, unshifted frequency obtained is \mbox{$1{,}819{,}599{,}021{,}549 \pm 9$ Hz}. The fractional statistical uncertainty of $5.5\times 10^{-12}$ is limited mainly by the instability of the atomic clock (Stanford Research Systems model FS725) used as a reference for stabilizing the repetition rate of the femtosecond frequency comb. Repeated frequency measurements performed over the course of five months are presented in figure \ref{Fig3}(b). A weighted fit of these four data points gives a center frequency of \mbox{$\bar{\nu}_D = 1{,}819{,}599{,}021{,}555 \pm 8$ Hz}.
 
The most significant systematics on the transition frequency are listed in Table \ref{Tab1}. Since the linear Zeeman shift and the frequency comb AC-Stark shift cancel by the chosen measurement scheme, the largest remaining frequency shift is due to the quadratic Zeeman shift, calculated to be \mbox{21.94(2) Hz} at a mean magnetic field of \mbox{6.500(3) G}. The electric-quadrupole shift from the gradient of the DC-trapping fields is \mbox{-0.79(2) Hz}. The only relevant light shift is due to the residual light of the AC-Stark-shift laser when switched off during spectroscopy by two acousto-optic modulators (AOM). To estimate this shift, we measured the fine structure level separation for two opposite detunings of the AC-Stark-shift laser and for ten times higher intensity. The measured shift, zero within our statistical uncertainty, bounds this systematic to \mbox{-0.3(1.0) Hz}. Differential AC-Stark shifts induced by the 729-nm and 397-nm beams are estimated to be well below the Hz level. The two repumper beams at 866 nm and 854 nm are physically blocked during spectroscopy and do not induce any additional bias. Since the polarizabilities of the $D_{5/2}$ and $D_{3/2}$ states differ by less than 1\% \cite{Safronova_2011}, the differential DC-Stark shift due to black-body radiation amounts to only 2(6) mHz. Stark shifts and second order Doppler effects due to residual micromotion of the ion can be estimated below the mHz level \cite{Berkeland_1998}. Hence, the major source of error originates from our GPS-disciplined rubidium standard, whose fractional inaccuracy was measured against an acetylene-stabilized ultra-stable fiber laser (\textit{Stabi$\lambda$aser} from Denmark's National Metrology Institute \cite{ stabilaser, Talvard2017}) to be $5\times 10^{-12}$. Correcting for the aforementioned systematics, our measurement corresponds to a fine structure level separation of \mbox{$\nu_D^{cor.} = 1{,}819{,}599{,}021{,}534 \pm 8$ Hz}. This result is consistent with a previous experiment where two phase-locked CW lasers were applied (\mbox{$1{,}819{,}599{,}021{,}504 \pm 37$ Hz}) \citep{Yamazaki2008}; however, we achieve nearly five times better accuracy.

\begin{table}
\caption{\label{Tab1}The systematic frequency shifts and their associated 1$\sigma$ standard errors in Hz.}
\begin{ruledtabular}
\begin{tabular}{lcc}
Effect & Shift (Hz) & Error (Hz) \\
\hline
2nd-order Zeeman & 21.94 & 0.02\\
Electric-quadrupole & -0.79 & 0.02\\
AC-Stark shifts: \\
AC-Stark-shift laser & -0.3 & 1.0\\
Laser at 729 nm & 0 & {\raisebox{0.6pt}\textless}0.2\\
Laser at 397 nm & 0 & {\raisebox{0.75pt}\textless}0.001\\
Lasers at 866 nm and 854 nm & 0 & 0\\
Black-body Radiation & 0.002 & 0.006\\
Excess micromotion & 0 & {\raisebox{0.75pt}\textless}0.001\\
Rb Standard & 0 & 9 \\ 
\\
Total & 20.9 & 9 \\
\end{tabular}
\end{ruledtabular}
\end{table}

\begin{figure}[t]
\includegraphics[width=0.48\textwidth]{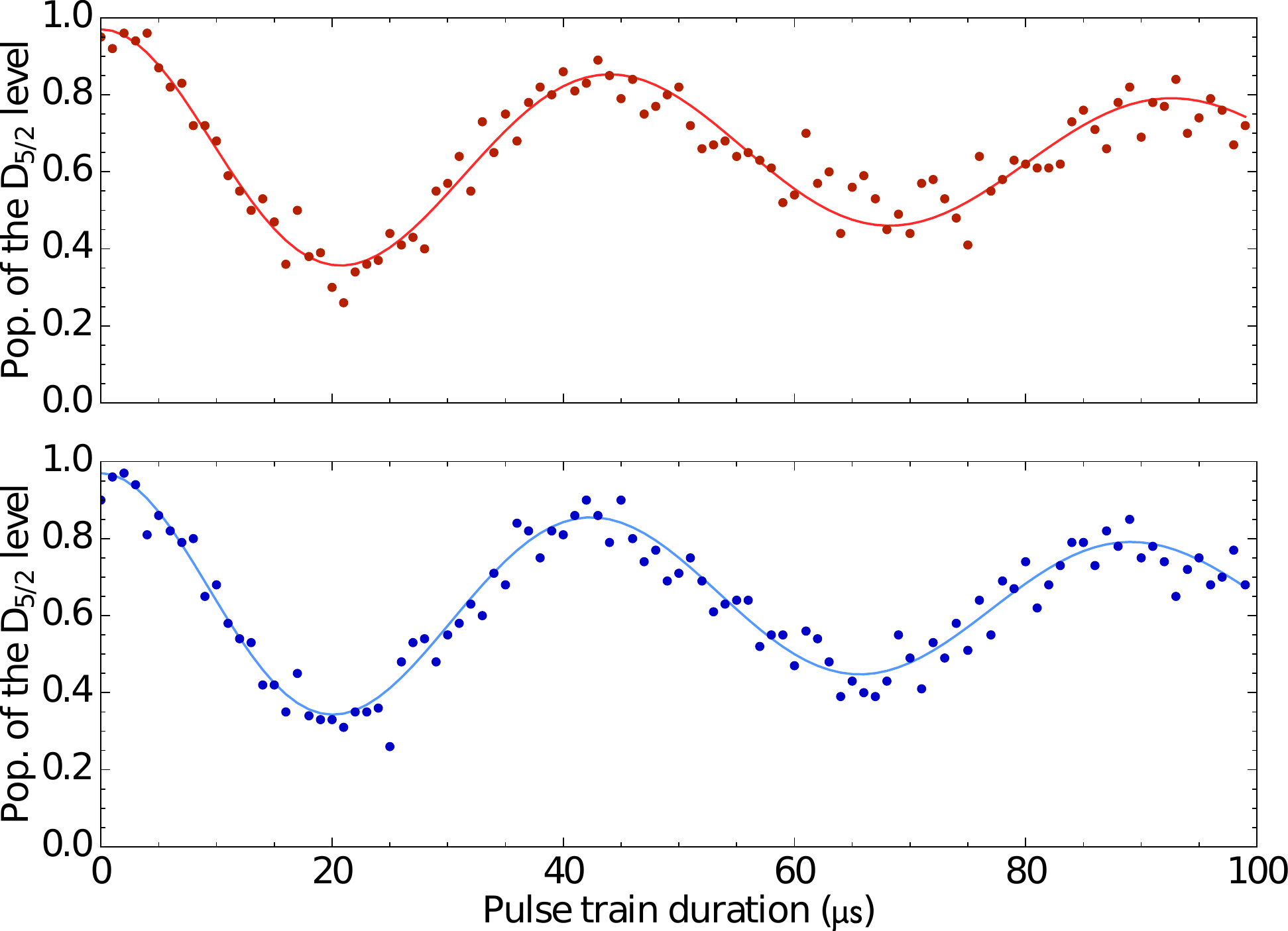}
\caption{Top (bottom): Evolution of the population of the $|D_{5/2}\rangle$ state versus duration of the Ti:Sapphire frequency comb's pulse train without (with) GDD compensation. Lines are fit to the experimental data \cite{supmat}.}
\label{Fig4}
\end{figure}

For the above high-precision spectroscopy, the differential AC-Stark shift $\delta\nu^{AC}_D$ essentially vanished due to the existence of a ``magic polarization'' and dispersion compensation of the comb's spectra was not really an issue, even though the effective Raman Rabi frequencies were roughly ten times lower than the maximum possible \cite{supmat}. For high-precision Raman spectroscopy in general, keeping the differential AC-Stark shift as small as possible for a given Rabi frequency is essential though, and group delay dispersion (GDD) compensation is necessary. This is as well the case for maximizing the attainable Rabi frequency for driving very far off-resonant Raman transitions (e.g. in molecular ions) or optimizing fast gate operations between Raman qubit states. 
To look into this aspect, we replaced our fiber femtosecond frequency comb by a mode-locked Ti:Sapphire solid-state laser from Coherent Inc. (model Mira 900 \cite{mira}). This laser not only provides three times more power, but its pulses (duration of 63 fs) at the output coupler are essentially Fourier-limited, hence allowing for a rather simple GDD compensation at the ion's position. To quantify the impact of GDD, one must sum over all the comb's frequency components and over the different intermediate Zeeman sub-levels $|i\rangle$ of the $P_{3/2}$ state, which, after adiabatic elimination of the intermediate levels and in the rotating wave approximation, results in the Raman Rabi frequency:
\begin{align}\label{RabiCoherent}
\Omega_R &=\eta|\sum\limits_{n,i} \frac{|\Omega^{g,i}_n \Omega^{e,i}_{n-q}|}{2\Delta_{n,i}}e^{i\psi^{e,g}_i}e^{i\delta\phi_n}| \\ &\equiv \eta_{eff}|\sum\limits_{n,i} \frac{|\Omega^{g,i}_n \Omega^{e,i}_{n-q}|}{2\Delta_{n,i}}e^{i\psi^{e,g}_i}| \nonumber
\end{align}
where $\Omega^{j,i}_n$ is the one-photon Rabi frequency of the tooth $n$ addressing the $|j=g,e\rangle$ to $|i\rangle$ transition, $\Delta_{n,i}$ is the Raman detuning of the Raman pair $(n-q,n)$ to the intermediate level $|i\rangle$, $\delta\phi_n$ is the phase difference between the two teeth of the Raman pair $(n-q,n)$, $\psi^{e,g}_i$ is the phase corresponding to the Raman path involving the intermediate level $|i\rangle$ and its associated Clebsch-Gordan coefficients \cite{supmat}, and $\eta$ ($\eta_{eff}$) is a measure of the efficiency, without (with) taking GDD into account, with which the total comb intensity is used. In the case of non-zero GDD, the spectral phase of the frequency comb to second order reads:
\begin{equation}\label{GDD}
\phi(\omega)=\phi_0+\tau_g(\omega-\omega_c)+\frac{D_2}{2}(\omega-\omega_c)^2
\end{equation}
where $\tau_g$ is the group delay and $D_2$ the GDD. As a result, \mbox{$\frac{\partial\delta\phi_n}{\partial n} = D_2q\omega_r^2\neq 0 $}, and the different Raman pairs interfere partially destructively, resulting in a Rabi frequency reduced by a factor $\eta_{eff}/\eta$. 
Raman Rabi oscillations obtained without (with) GDD compensation are shown in figure \ref{Fig4} top (bottom). As seen, the Raman Rabi frequencies are nearly the same ($\approx 21$ kHz). However, the two measurements were carried out at different comb intensities: 47(1) W/mm$^2$ (uncompensated case) and 38(1) W/mm$^2$ (compensated case). The lines are fits to the experimental data with two adjustable parameters: $\eta_{eff}$ and the effective linewidth of the Mira frequency comb \mbox{($\Delta\nu_{eff} = 43(2) $ kHz)} \cite{supmat}. Remarkably, $\eta_{eff}$ is found to be 0.72(3) (top curve) and 0.92(3) (bottom curve), the latter corresponding to a 92(3)\% use of the total comb power when GDD is compensated. For the uncompensated case, GDD was measured by frequency resolved optical gating using a ``GRENOUILLE'' device to be $D_2 = 2600$ fs$^2$ \cite{supmat}. Taking this value into account in Eq.\ref{RabiCoherent}, we obtain the same scaling factor $\eta = 0.92(3)\approx 1$ as for the compensated case (for which $\eta=\eta_{eff}$), hence validating our simple model and demonstrating that we are able to predict the dynamics of frequency-comb-driven Raman transitions based on the comb's spectral properties.

In summary, we have demonstrated direct frequency-comb-driven stimulated Raman transitions between the 1.8 THz separated $3d~^2D_{3/2}$ and $3d~^2D_{5/2}$ fine structure states of a single trapped $^{40}$Ca$^+$ ion. We achieved high-contrast Rabi oscillations and high-resolution spectroscopy with a fiber-based frequency comb, and proved, using a solid-state Ti-sapphire frequency comb, that all the individual comb teeth can contribute coherently to the effective Rabi frequency if group delay dispersion is fully compensated. Remarkably, we used this technique to improve the knowledge of the absolute transition frequency by nearly a factor of five. With a better atomic clock reference and a better locking scheme of the frequency comb, much higher resolution is within reach. Such high resolution spectroscopy performed on other calcium isotopes could, e.g., in combination with precise measurements of the $4s~^2S_{1/2}-3d~^2D_{5/2}$ transition, improve bounds on new physics beyond the standard model \cite{Berengut2018}. Moreover, this versatile technique, which takes advantage of the full tunability of femtosecond frequency combs, can be generalized to drive sub-repetition rate transitions by splitting the comb's beam into two AOM-controlled beams in a similar way to \cite{Hayes2010}, hence allowing for driving any Raman transition between state-separations ranging from the sub-kHz range to a few tens of THz. The method could therefore be used to coherently manipulate anything from, e.g., hyperfine-resolved rovibrational transitions in molecules \cite{Staanum2010,Hansen2014,Wolf2016,Chou2017a} to hyperfine transitions in highly charged ions \cite{Gillaspy2001,Schmoger2015}, and eventually open up paths towards new qubit systems for quantum technology.

\begin{acknowledgments}
We acknowledge the Danish national laser infrastructure, LASERLAB.DK, established through the support of the Danish Ministry of Research and Education, for financial support and access to the fiber-based frequency comb. We also acknowledge support from Innovation Fund Denmark, through the Quantum Innovation Center, Qubiz, for financial support and access to the \textit{Stabi$\lambda$aser}. C.S. and M.V.D. acknowledge supports from LASERLAB.DK. S.M., K.F. and M.V.D acknowledge support from the European Commission through the Marie Curie Initial Training Network COMIQ (grant agreement no 607491) under FP7. M. D. acknowledges support from the Danish Council of Independent Research through the Sapere Aude Advanced grant. We are extremely grateful to H. Stapelfeldt from the Department of Chemistry, Aarhus University for lending us their GRENOUILLE device.
\end{acknowledgments}

\bibliography{CombTechnique}

\onecolumngrid

\section{Magic Polarization} 

\begin{figure}[ht]
\includegraphics[width=0.95\textwidth]{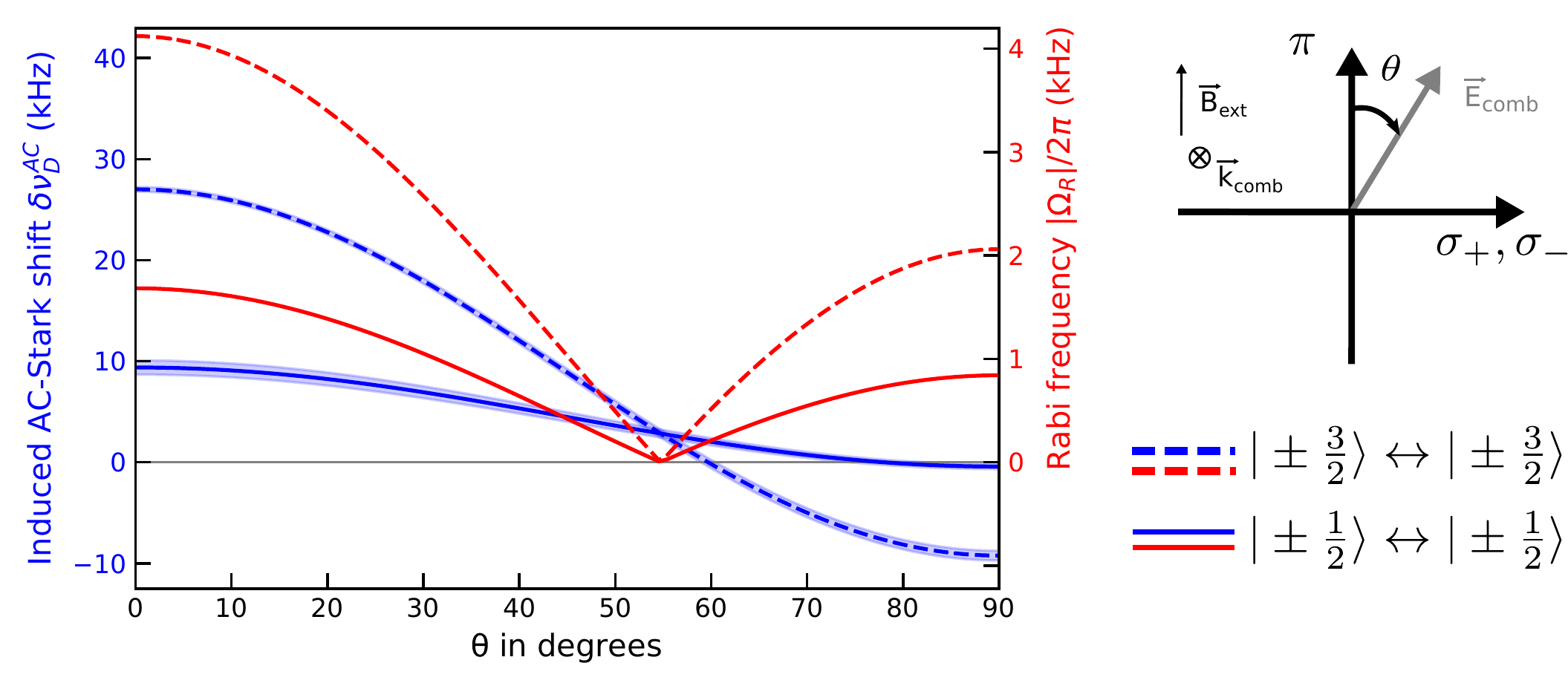}
\caption{Solid and dashed blue lines correspond to the differential AC-Stark shifts induced by the fiber-based femtosecond frequency comb laser on the $|\frac{3}{2},\pm \frac{1}{2}\rangle \rightarrow |\frac{5}{2},\pm \frac{1}{2}\rangle$ and $|\frac{3}{2},\pm \frac{3}{2}\rangle \rightarrow |\frac{5}{2},\pm \frac{3}{2}\rangle$ transitions respectively. Shaded areas represent the uncertainty on the calculated curves due to the uncertainty of the Einstein A coefficients. Solid and dashed red lines represent the corresponding Raman Rabi frequencies calculated from Eq.(\ref{Sup_RabiCoherent}) and with $\eta_{eff}=0.1$.}
\label{SFig1}
\end{figure}

Off-resonant coupling of the femtosecond frequency comb light to the $P_{1/2}$ and $P_{3/2}$ states induces AC-Stark shifts on both the $D_{3/2}$ and $D_{5/2}$ states which read:
\begin{equation}\label{Sup_ACStark}
\Delta^{AC}_g=\sum\limits_{n,i} \frac{|\Omega^{g,i}_n|^2}{4\Delta_{n,gi}}+\frac{|\Omega^{g,i}_n|^2}{4\Delta^+_{n,gi}}
\end{equation}  
where $\Omega^{g,i}_n$ is the coupling strength of the tooth $n$ to the $g\leftrightarrow i$ transition and depends on the dipole matrix elements, the polarization of the frequency comb and on the tooth's intensity \cite{James1998}. Both $\Delta^+_{n,gi}=n\omega_r+\omega_{CEO}+\omega_{gi}$ (where $\omega_{CEO}/2\pi$ is the carrier envelope offset frequency) and the detuning $\Delta_{n,gi}=n\omega_r+\omega_{CEO}-\omega_{gi}$ are much larger than the Zeeman splitting and are assumed to be the same for all Zeeman sublevels. The differential AC-Stark shift on the D-fine-structure splitting $\delta\nu_D^{AC}=(\Delta^{AC}_{D5/2}-\Delta^{AC}_{D3/2})/2\pi$ is plotted in figure \ref{SFig1} as a function of the angle $\theta$ between the polarization of the frequency comb $\vec{E}_{comb}$ and the external magnetic field $\vec{B}_{ext}$. Note that the linear polarized light of the frequency comb, with $\vec{k}_{comb}\perp \vec{B}_{ext}$, corresponds to $\pi$ and an equal superposition of $\sigma_+$ and $\sigma_-$ photons. The differential AC-Stark shift is consequently the same for symmetric transitions $|\frac{3}{2},\pm m_i\rangle \rightarrow |\frac{5}{2},\pm m_f\rangle$.

As shown by the blue dashed line in figure \ref{SFig1}, the differential AC-Stark shift cancels out
for the $|\frac{3}{2},\pm \frac{3}{2}\rangle \rightarrow |\frac{5}{2},\pm \frac{3}{2}\rangle$ transition when the linear polarization is rotated by $\theta=59(1)^{\circ}$, as well as for the $|\frac{3}{2},\pm \frac{1}{2}\rangle \rightarrow |\frac{5}{2},\pm \frac{1}{2}\rangle$ transition when $\theta=79(4)^{\circ}$ (blue solid line). The discrepancy of the latter from the experimentally determined magic polarization ($\theta=88(1)^{\circ}$) could be explained by off-resonant coupling of the frequency comb light to higher electronic energy levels, as estimated from the measured comb intensity and calculated polarizabilities of the $3d~^2D_{3/2}$ and $3d~^2D_{5/2}$ energy levels \citep{Safronova_2011}. Note that being proportional to the combs intensity, these couplings (and associated shifts) do not impact the accuracy of the extrapolated transition frequency at zero intensity. 

Figure \ref{SFig1} also shows, for the same transitions, the dependence of the Raman Rabi frequencies with respect to $\theta$ (see red curves), calculated using the formula: 
\begin{equation}\label{Sup_RabiCoherent}
|\Omega_R|=\eta_{eff}|\sum\limits_{n,i} \frac{|\Omega^{g,i}_n \Omega^{e,i}_{n-q}|}{2\Delta_{n,i}}e^{i\psi^{e,g}_i}|
\end{equation}
where $\eta_{eff}$ is a prefactor with a value between 0 and 1, dependent on the degree of group delay dispersion (GDD) at the position of the ion (discussed in detail below). The phase $\psi^{e,g}_i$ corresponds to the Raman path involving the intermediate level $|i\rangle$ and its associated Clebsch-Gordan coefficients. This formula takes into account the interference between the different Raman paths, and as can be seen in figure \ref{SFig1}, the Raman Rabi frequencies of both transitions become zero for $\theta\approx55^{\circ}$. Note that around the angle corresponding to the magic polarization for the $|\frac{3}{2},\pm \frac{1}{2}\rangle \rightarrow |\frac{5}{2},\pm \frac{1}{2}\rangle$ transition, the Raman Rabi frequency is still about half of the maximum achievable value that occurs when the comb's polarization is parallel to the magnetic field axis.

\begin{figure}[ht]
\includegraphics[width=0.55\textwidth]{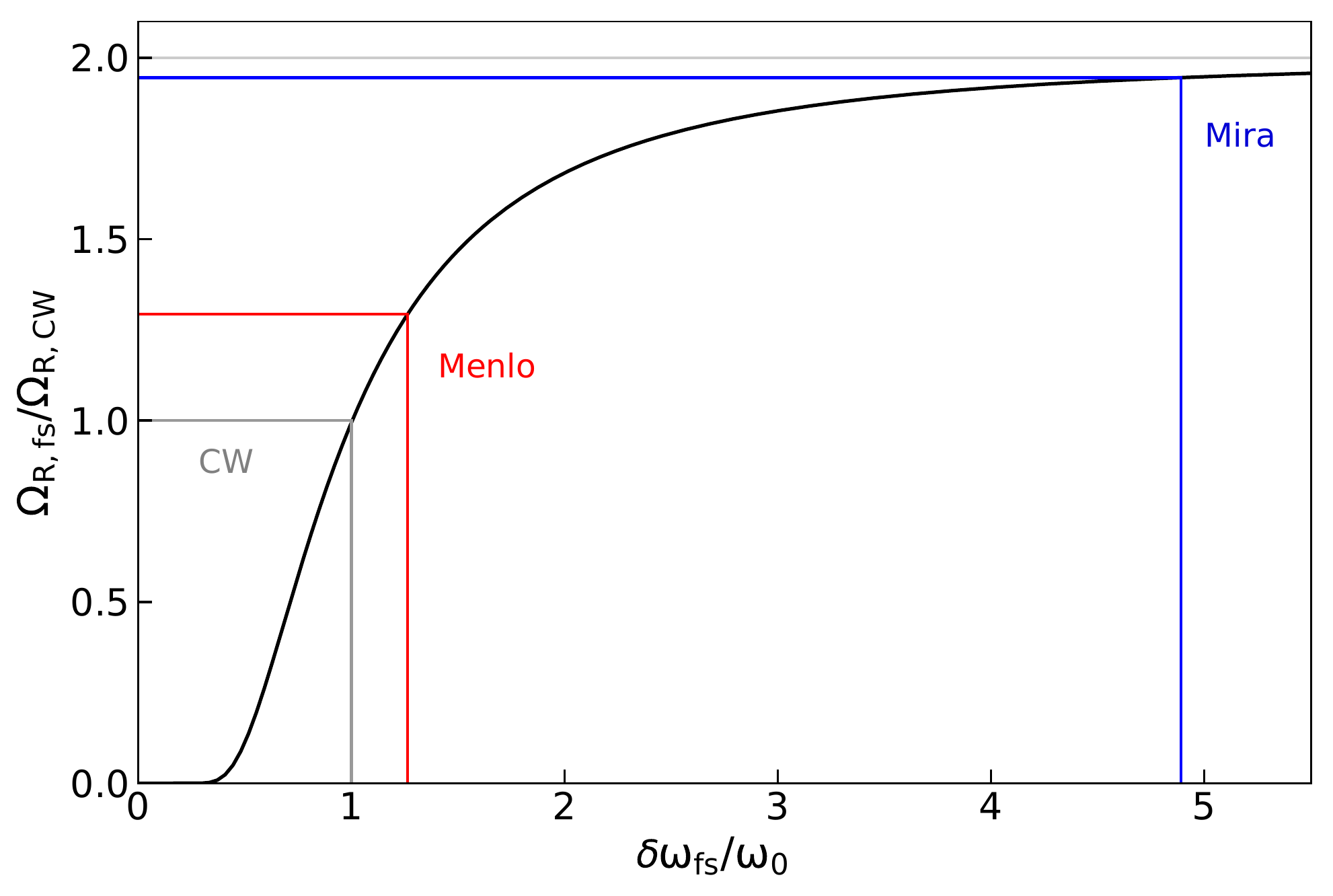}
\caption{Raman Rabi frequency versus spectral bandwidth of the femtosecond frequency comb. The curve is calculated from Eq.(\ref{Sup_RabiCoherent}) for a Gaussian spectral distribution of full-width-half-maximum $\delta\omega_{fs}$ and a given total intensity of the frequency comb $I_{fs}$. The curve is normalized by the Raman Rabi frequency calculated for two phase-locked CW lasers with total intensity $I_{CW}=I_{fs}$. $\omega_{0}$ is the transition frequency. The spectral bandwidths of the fiber-based frequency comb ($\delta\omega_{fs}/\omega_{0}\approx 1.3$) and of the solid-state Ti:Sapphire frequency comb ($\delta\omega_{fs}/\omega_{0}\approx 4.9$) are shown in red and blue respectively.}
\label{SFigbis}
\end{figure}

Figure \ref{SFigbis} shows the dependence of the Raman Rabi frequency on the spectral bandwidth of the femtosecond frequency comb, calculated from Eq.(\ref{Sup_RabiCoherent}) without taking group delay dispersion into account ($\eta_{eff}=1$) and for a constant average Raman detuning ($\Delta = 1/N\sum\Delta_n$). For a Gaussian spectral distribution of full-width-half-maximum $\delta\omega_{fs}$ equal to the transition frequency $\omega_{0}$, and for a total comb intensity $I_{fs}$, the Raman Rabi frequency is the same as for two phase-locked CW lasers with same total intensity ($I_{CW}=I_{fs}$). The Raman Rabi frequency for two CW lasers $\Omega_{R,CW}$ can be calculated using Eq.(\ref{Sup_RabiCoherent}) with two comb teeth of intensity $I_{fs}/2$ and a Raman detuning $\Delta$. For a spectral bandwidth significantly wider than the transition frequency, the Raman Rabi frequency is twice the Raman Rabi frequency for CW lasers ($\Omega_{R,fs} = 2\Omega_{R,CW}$). This is due to the fact that each comb tooth number $n$ contributes to two Raman processes driven by the pair of teeth ($n,n-q$) and ($n,n+q$). As shown in figure \ref{SFig2}, the spectrum of the Ti:Sapphire solid-state laser (Mira frequency comb) is very close to being a single Gaussian shape with $\delta\omega_{fs} \approx 4.9 \omega_0$, and with almost each comb tooth contributing to two Raman processes, the obtained Raman Rabi frequency is close to being $2\Omega_{R,CW}$. On the other hand, the spectrum of the fiber-based frequency comb (Menlo) is much more complex and also narrower (see figure \ref{SFig2}). However, taking the full spectrum into account instead of the main Gaussian feature centered around 789 nm and using the same total intensity, changes the calculated Raman Rabi frequency by less than 1\%. Note that despite this relatively narrow spectral bandwidth of only $\approx 1.3 \omega_0$, the achieved Raman Rabi frequency is larger than $\Omega_{R,CW}$.

\section{Group delay dispersion}

\begin{figure}[ht]
\includegraphics[width=0.6\textwidth]{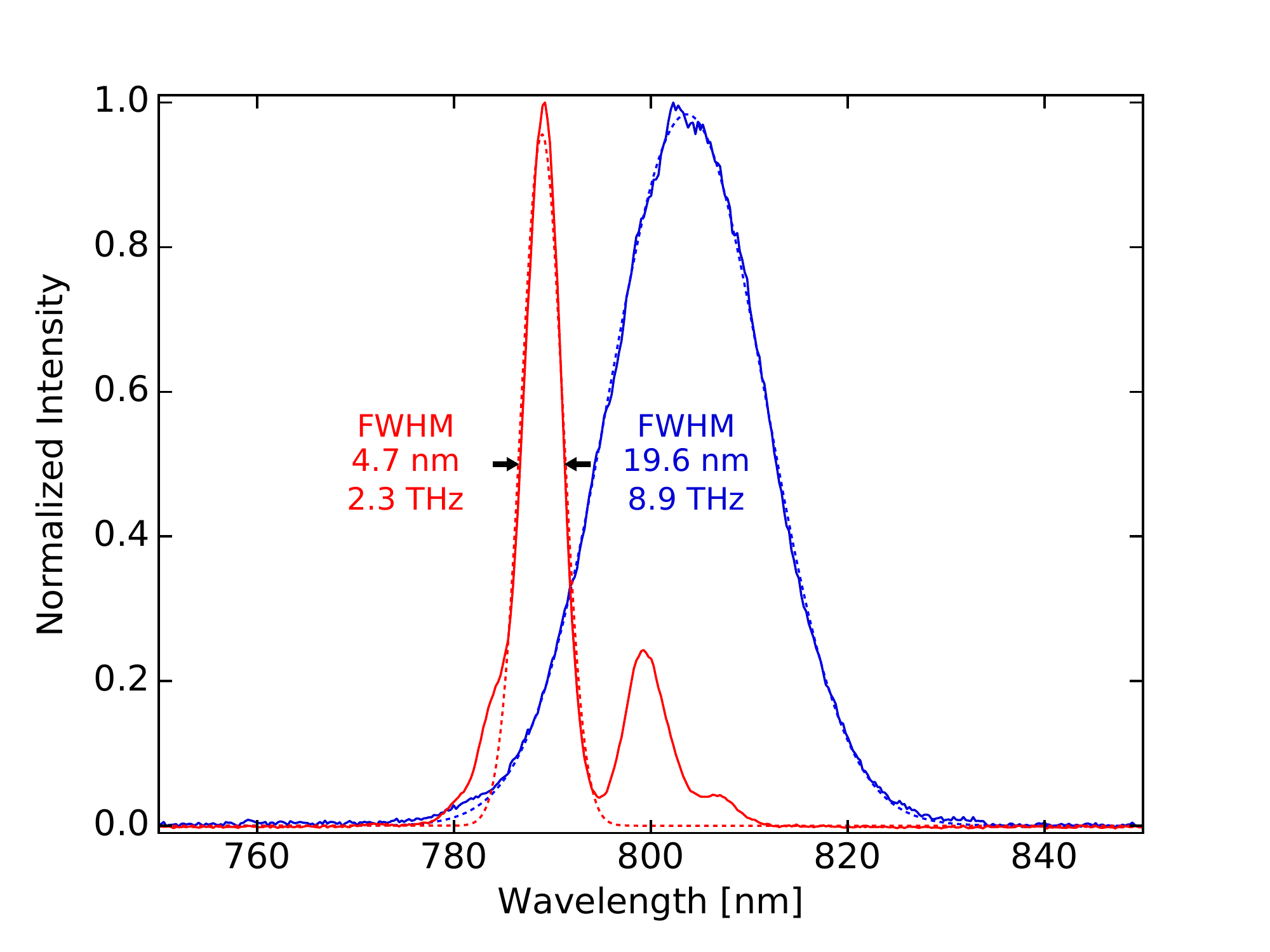}
\caption{Spectral distributions of the two frequency comb applied in the experiments. The solid red curve represents the spectrum of the fiber-based commercial femtosecond frequency comb, while the solid blue curve represents the spectrum of the solid-state Ti:Sapphire femtosecond frequency comb. The dashed curves are Gaussian fits to the main features in the two spectra.}
\label{SFig2}
\end{figure}

The very different spectral distributions of the two frequency combs applied in the experiments are shown in figure~\ref{SFig2}. While the spectrum of the Mira (Ti:sapphire solid-state laser) is very close to being a single Gaussian shape with a pulse length at the output coupler of the laser of 63~fs, and hence close to being bandwidth-limited, the MenloSystems laser (fiber-based laser) spectrum is much more complex due to the non-linear processes in the doubling and amplifying stages after the seed laser at 1.6~$\mu$m. As a consequence, the pulse length of the latter is about 400~fs, and far from being bandwidth-limited. Hence, there is no simple phase relation between the comb teeth at the output facet of the fiber laser, which makes it challenging to provide dispersion-free fs pulses at the position of the ion. The effective Raman Rabi frequency achieved with this laser is therefore reduced by a factor $\eta_{eff}\approx 0.1$. In contrast, the pulses of the Ti:sapphire laser are, at the laser output coupler, essentially fully dispersion compensated. This laser consequently makes it easier to compensate for dispersion at the ion's position since GDD is introduced only by optical elements along the path.

\begin{figure}[ht]
\includegraphics[width=0.7\textwidth]{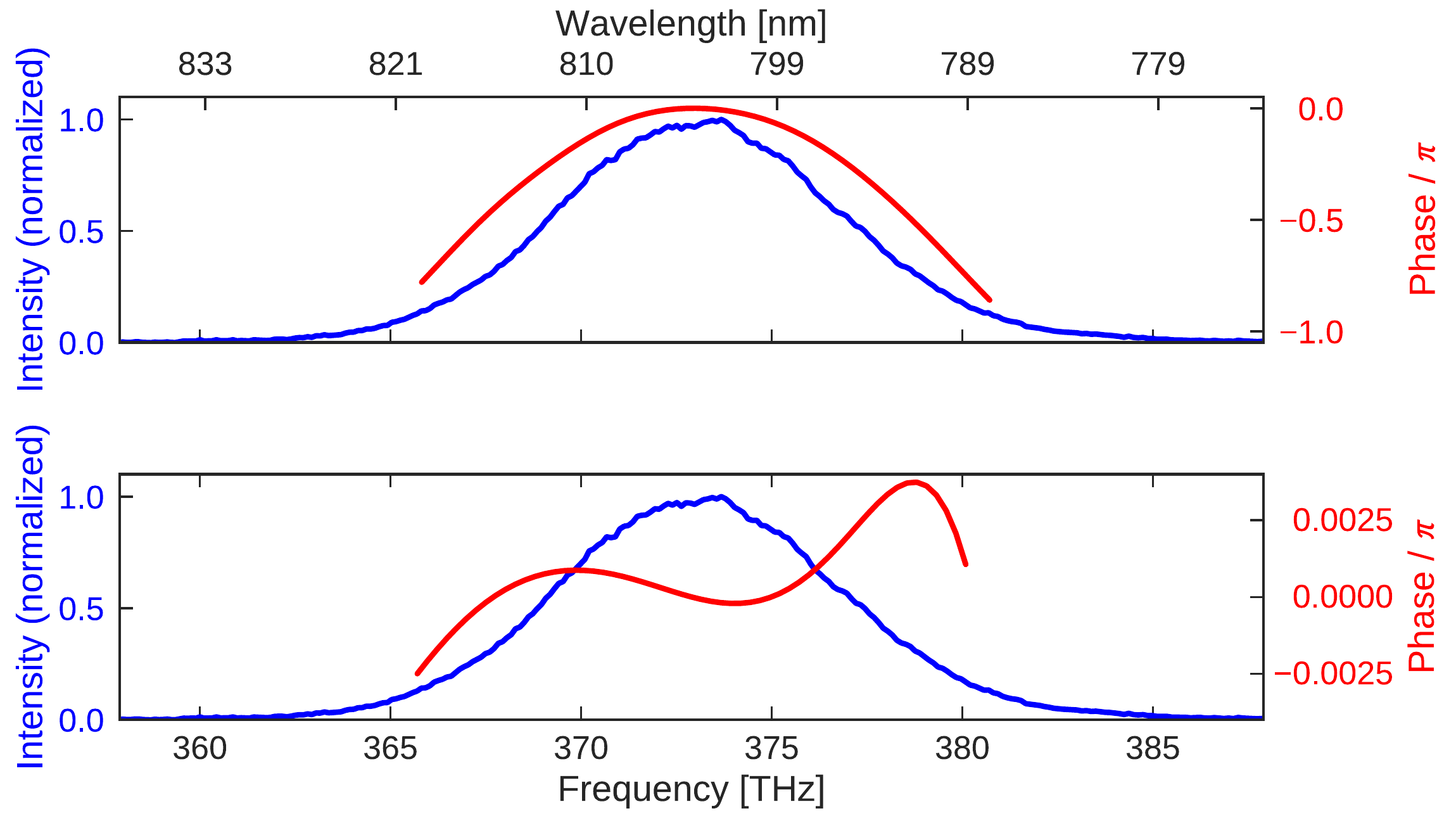}
\caption{Top (bottom): Spectral intensity (blue) and spectral phase (red) of the Ti:Sapphire frequency comb at a position equivalent to the one of the ion without (with) GDD compensation, measured respectively with an optical spectrum analyzer and a ``GRENOUILLE'' device for frequency resolved optical gating. The corresponding Rabi oscillations are shown in figure 4 of the main text.}
\label{SFig3}
\end{figure}

When propagating in dispersive materials (e.g. glass) the spectral phase of the frequency comb can be written:  
\begin{align*}\label{Sup_TaylorGDD}
\phi(\omega)&=\phi(\omega_c)+\frac{\partial\phi}{\partial\omega}\Bigr|_{\omega=\omega_c}(\omega-\omega_c)+\frac{1}{2}\frac{\partial^2\phi}{\partial\omega^2}\Bigr|_{\omega=\omega_c}(\omega-\omega_c)^2+\mathcal{O}(\omega^3)\\
&=\phi(\omega_c)+\tau_g(\omega-\omega_c)+\frac{D_2}{2}(\omega-\omega_c)^2+\mathcal{O}(\omega^3)
\end{align*}
where the group delay $\tau_g$ and group delay dispersion $D_2$ are characteristics of the dispersive material. Neglecting higher order terms, the spectral phase of the femtosecond frequency comb is quadratic due to GDD. A measurement of this quadratic dependence at a position in the beampath equivalent to the one of the ion is shown in figure \ref{SFig3} (top) for the Ti:Sapphire femtosecond laser. Using a ``GRENOUILLE'' device for frequency resolved optical gating we obtain $D_2=2600$ fs$^2$. This value, which was also consistently checked with the amount of dispersive material the femtosecond frequency comb beam is going through, reduces the effective Raman Rabi frequency by a factor $\eta_{eff}\approx 0.72(3)$. Figure \ref{SFig3} (bottom) shows the frequency comb spectral phase at the same position when GDD is compensated. The residual phase variation ($\approx 5\pi$ mrad) is due to higher order terms (mostly third order dispersion) in the Taylor expansion which cannot be compensated using a simple prism pulse compressor. In this case, as determined from figure 4(b) of the main text, $\eta_{eff}\approx 0.92(3)$ and nearly all the frequency components contribute to drive the Raman transition, hence allowing for the optimal use of the total femtosecond frequency comb's power.   

\section{Evolution dynamics of the populations}

For the special case of a repetition rate slowly varying on the order of the time between two measurement realizations ($\sim 100$ ms), the statistical average for the whole experiment can be modeled as a sum of Rabi flopping traces with different Raman detunings $\delta=q\omega_r-\nu_D$ and distributed according to a chosen distribution function $g(\delta)$ such that the dynamics of the $D_{5/2}$ state population can be written as:
\begin{equation}\label{Sup_LinewidthModel}
P_{D_{5/2}}(t)=\int^{+\infty}_{-\infty} d\delta~ g(\delta) \frac{\Omega_R^2}{\Omega_R^2+\delta^2}\Big[1-\cos(\sqrt{\Omega_R^2+\delta^2}t)\Big]/2
\end{equation}
where $\Omega_R$ is given by Eq.(\ref{Sup_RabiCoherent}) and where in our case the effective Raman lineshape $g(\delta)$ is Gaussian. Fitting the population dynamics of the $D_{5/2}$ state obtained with the Ti:Sapphire frequency comb (see Fig.\ref{Fig4}) using Eq.(\ref{Sup_LinewidthModel}) gives a full-width-at-half-maximum of $\Delta\nu_{eff} = 43(2)$ kHz. This effective linewidth, which also fits the Gaussian spectroscopy profiles we obtain with the Ti:Sapphire frequency comb, does not allow for reaching the 8~Hz accuracy we achieve with the fiber-based frequency comb in a reasonable experimental time. Note that this effective linewidth corresponds to a repetition rate instability of $\Delta\nu_r = \Delta\nu_{eff}/q = 1.8(1)$ Hz (compared to $\Delta\nu_r \approx 0.1$ Hz for the fiber-based frequency comb). Whereas such a stability would be sufficient to, e.g., drive GHz transitions with full transfer efficiency, it is not for driving THz transitions (see Fig.\ref{Fig4}) which requires higher stability of the repetition rate. This can be achieved by replacing the piezoelectric actuator and cavity's end mirror used for stabilizing the cavity length of the Ti:Sapphire frequency comb.

\end{document}